\documentclass[11pt]{article}
\usepackage{moriond,epsfig}

\title{DIRECT PHOTONS IN p+p, d+Au AND Au+Au COLLISIONS AT
  $\sqrt{s_{NN}}$ = 200 GeV}
\author{B. SAHLMUELLER for the PHENIX Collaboration}
\address{University of M\"unster, Institut f\"{u}r Kernphysik, Wilhelm-Klemm-Str. 9, 48149 M\"{u}nster, Germany}
\begin{document}
\vspace*{4cm}
\maketitle
\abstracts{The PHENIX experiment has measured direct photons at $\sqrt{s_{NN}}$ =
200 GeV in $p+p$, $d$+Au and Au+Au collisions. For $p_{T}$ $<$ 4 GeV/$c$, the
internal conversion into $e^{+}e^{-}$ pairs has been used to measure the
direct photons in Au+Au.}

\section{Introduction}
Direct photons are a unique probe to study the hot and dense matter
produced in heavy ion collisions at RHIC. They provide information
about the thermalized state of such collisions. Moreover, high-$p_{T}$
direct photons can be used as a measure of the rate of initial hard
parton-parton scatterings.\\
In $p+p$ collisions, direct photons are produced by hard scattering processes. A direct
photon measurement thus is an important baseline for the understanding of the
hard scattering contribution in Au+Au collisions as well as - in its
own right - a good test of pQCD calculations. Measurements
in $d$+Au collisions allow a quantification of possible initial state
effects. In (central) heavy ion collisions, the temperature of the
collision can be obtained via thermal direct photons from the QGP that are expected
to be the dominant source of direct photons at 1 GeV/$c < p_{T} <$ 3
GeV/$c$~\cite{Turbide2004} while at high transverse momenta hard direct photons can help
understanding the observed hadron suppression~\cite{Adler2003a}. There are also
theoretical predictions for direct photons produced in the interaction
of a jet and the created medium~\cite{Turbide2005}.

\section{Measurement of Direct Photons}
\begin{figure}[t]

\begin{minipage}[t]{65mm}
\epsfig{file=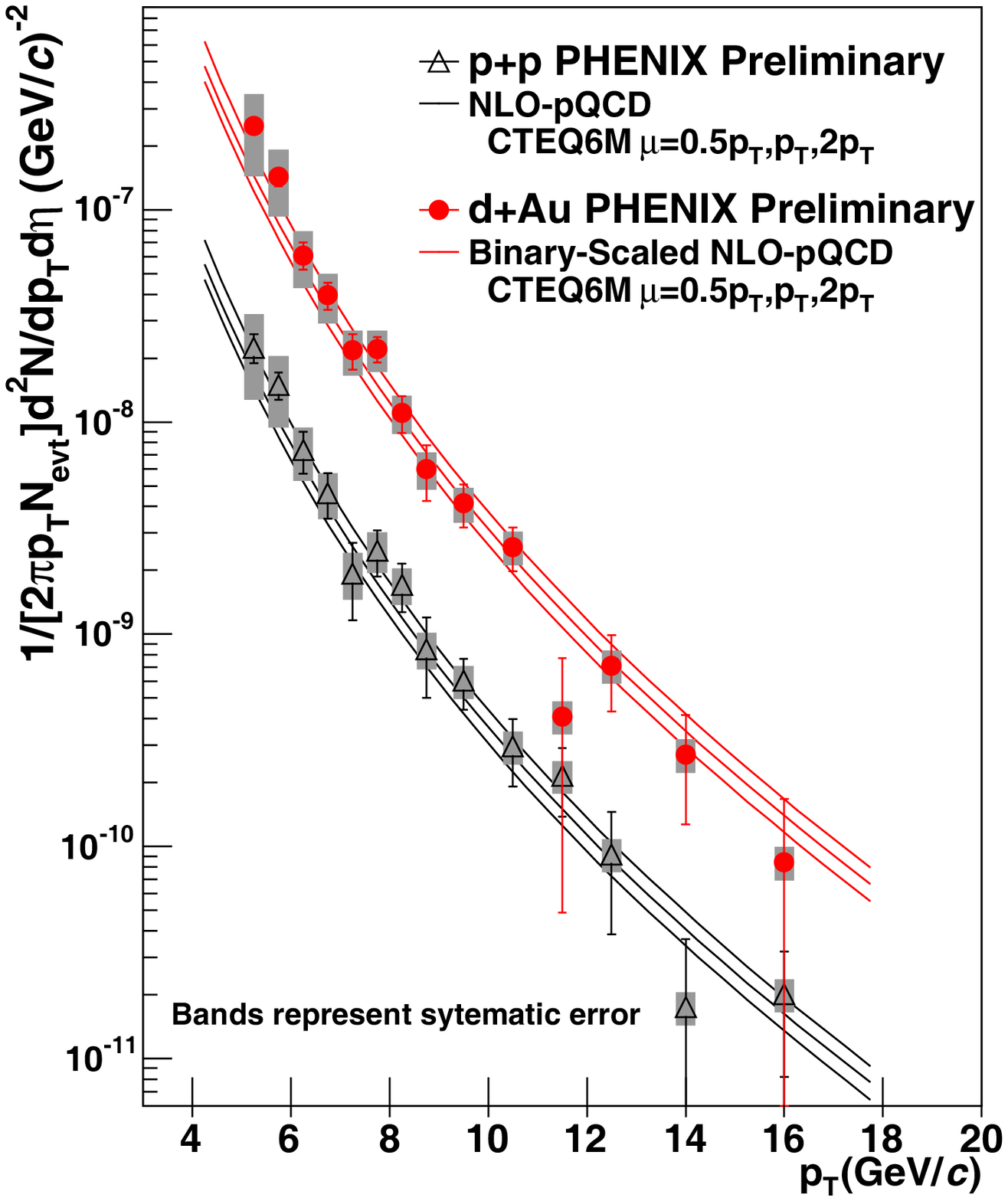,height=3.25in}
\caption{\label{fig:pp_dau}Preliminary direct photon invariant yield
  as a function of $p_{T}$ as measured in $p+p$ and $d+$Au at
  $\sqrt{s_{NN}}$ = 200 GeV. The solid curves are pQCD predictions for
  three different scales~\protect\cite{Vogelsang}. }
\end{minipage}
\hspace{\fill}
\begin{minipage}[t]{90mm}
\epsfig{file=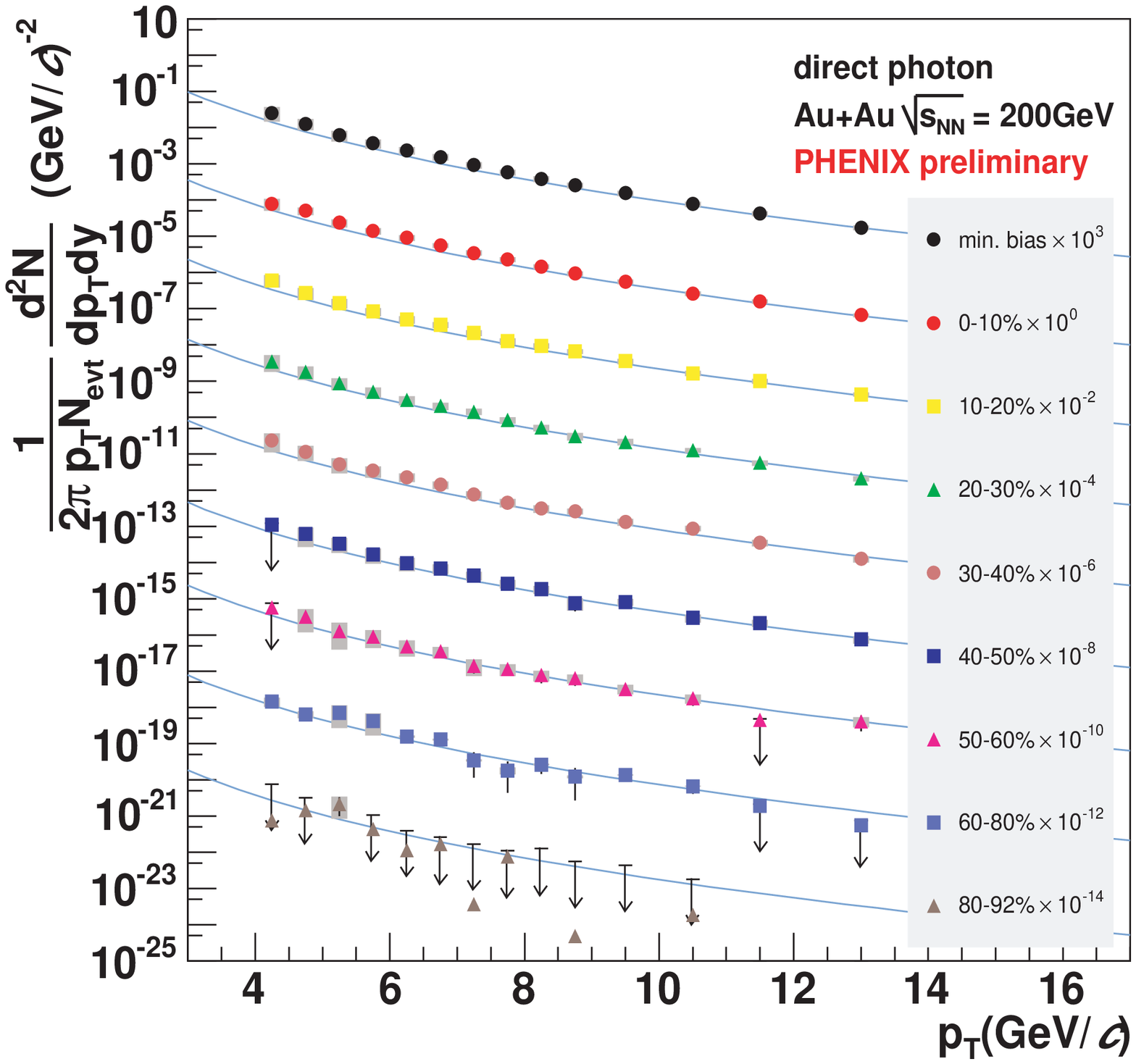,height=3.3in}
\caption{\label{fig:auau}Preliminary direct photon invariant yield as
 a function of $p_{T}$ as measured in Au+Au at $\sqrt{s_{NN}}$ = 200
 GeV for different centrality classes and minimum bias. The solid
 curves are pQCD predictions~\protect\cite{Vogelsang}, scaled with the corresponding nuclear
 overlap function $T_{AB}$. }
\end{minipage}

\end{figure}
The measurement is challenging due to a large background from decaying
hadrons such as the $\pi^{0}$ and the $\eta$. Two methods have been applied to
measure the direct photon signal: The
so-called subtraction or cocktail method  has been used for all
collision systems. The idea of this method is to subtract the photons
from hadronic decays from all inclusive photons; it has been well
established in different
analyses.~\cite{Adler2005a}$^{,}$~\cite{Okada2005}. This method has been
applied to the 2003 $p+p$ and
$d$+Au dataset as well as to the large 2004 Au+Au dataset with about 10
times more events than in the 2002 dataset. In the Au+Au case, direct
photons have also been measured via their internal conversion into
$e^{+}e^{-}$ pairs. This measurement has already been described elsewhere
\cite{Bathe2005} and has lead to a significant improvement of the
measurement at lower transverse momenta.\\

The idea of the internal conversion method is to use a process
corresponding to the $\pi^{0}$ (or $\eta$) Dalitz decay where one
decay photon is a virtual photon decaying into an $e^{+}e^{-}$
pair. The invariant mass distribution of these electron-positron pairs
is given by~\cite{KrollWada}\\
\begin{equation}
  \frac{1}{N_\gamma} \frac{dN_{ee}}{dm_{ee}} = \frac{2\alpha}{3\pi}
   \sqrt{1 - \frac{4m_e^2}{m_{ee}^2}} \left(1 + \frac{2m_e^2}{m_{ee}^2}\right) 
   \frac{1}{m_{ee}} \mid F(m_{ee}^2) \mid^2 \left(1 - \frac{m_{ee}^2}{M^2}\right)^3 \; .
 \label{eq:KW}
\end{equation}
$\;$\\
The same formula applies to any source of real photons as each such
source also produces virtual photons at very low masses. In the case
of direct photons, there is no phase space limitation when $m_{ee} <<
p_{T}^{photon}$.\\

The great advantage of this method is, that the $e^{+}e^{-}$ pairs
from the $\pi^{0}$ Dalitz decay are suppressed at higher invariant
masses due to the limited phase space. By measuring the pair yield in
such a mass region, e.g. for $90 < m_{ee} < 300$ GeV/$c^{2}$, the
decay photon background can be mostly eliminated. As
$\gamma_{direct}^{*}/\gamma_{incl.}^{*} =
\gamma_{direct}/\gamma_{incl.}$ must be satisfied to derive real
photons from the measured virtual photons, one has to relate the
obtained yield to the yield in a region with an unrestricted phase
space. The term $R_{\gamma_{data}} = N_{data}(90-300$ MeV$)/N_{data}(0-30$ MeV$)$ is the ratio of
the measured yields in the two intervals, it is used to calculate
$\gamma_{direct}^{*}/\gamma_{incl.}^{*}$. Therefore, the ratios
$R_{\gamma^{*}_{direct}}$ and $R_{\gamma^{*}_{hadron}}$ are precisely calculated
from Equation \ref{eq:KW} that describes the form of the corresponding  electron-positron pair spectra. With the knowledge of these ratios, the ratio of the
virtual direct photons and the virtual inclusive photons can be
calculated as $\frac{\gamma_{direct}^{*}}{\gamma_{incl.}^{*}} =
\frac{N^{0-30}_{\gamma_{direct}^{*}}}{N^{0-30}_{\gamma_{incl.}^{*}}} =
\frac{R_{data} - R_{hadron}}{R_{direct} - R_{hadron}}$. To get the
final direct photon yield, the calculated $\gamma_{direct}/\gamma_{incl.}$ is applied to the real inclusive photon yield from the measurement with the subtraction method~\cite{Adler2005a}.\\

The uncertainty of the internal conversion method is much smaller than
the uncertainty of the classical method as most errors cancel in the
ratio. Therefore, the systematic error is dominated by the 20 \%
uncertainty
of the
$\eta/\pi^{0}$-ratio~\cite{Adler2005b} that translates to the same 
uncertainty in the direct photon yield. Other contributions to the
total systematic error of 25 \% are the error of the inclusive photon
yield (10 \%) and the acceptance for $e^{+}e^{-}$ pairs (5 \%).

\section{Results}

\begin{figure}
\epsfig{file=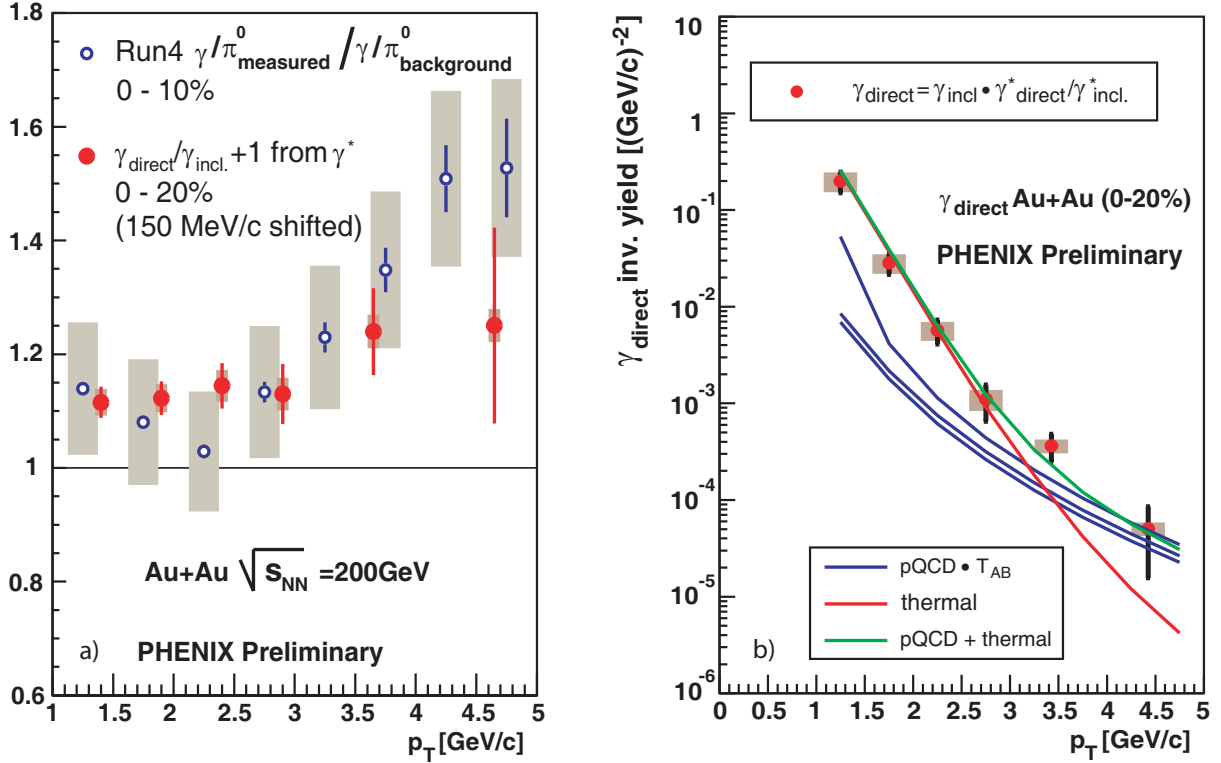,width=160mm}
\caption{\label{fig:auau_thermal}
a$)$ Direct photon excess for the most central events above the decay
photon background for the subtraction and the virtual photon
method in Au+Au collisions. b$)$ Direct photon yield for the same
centrality from the virtual photon measurement in comparison
with different theory
calculations: scaled pQCD~\protect\cite{Vogelsang}, thermal
model~\protect\cite{thermal} and the sum of both calculations.}
\end{figure}

The preliminary results for $p+p$ and minimum bias $d+$ Au collisions are shown in
Figure \ref{fig:pp_dau} and compared with a next-to-leading-order
perturbative-QCD (NLO pQCD) calculation~\cite{Vogelsang}. The data are
consistent with the calculation for the whole shown $p_{T}$ range
between 5 and 16 GeV/$c$. Therefore it is justified to use the pQCD
calculation as reference for hard direct photons, i.e. direct photons
at high transverse momenta, also in Au+Au collisions at the same
energy. Preliminary Au+Au results from recent measurements using the
large 2004 data sample, are shown for a $p_{T}$ region of 4 GeV/$c$ to
13 GeV/$c$ in Figure \ref{fig:auau} for eight
different centrality classes and for minimum bias. The data agree with
the pQCD expectation for all centralities. Compared to earlier
measurements~\cite{Adler2005a}, the errors are smaller and the
$p_{T}$ region has been extended. Further work is done to improve the
significance and to further extend the $p_{T}$ range.\\

For the lower $p_{T}$ region, the direct photon spectrum has also been
obtained with the internal conversion method. The significance of the
preliminary result is compared to the conventional subtraction method in
Figure \ref{fig:auau_thermal} a). The result from the subtraction method is shown as
the so-called double ratio
$(\gamma/\pi^{0}|_{meas})/(\gamma/\pi^{0}|_{background})$ while the
virtual photon result is shown as
$\gamma_{direct}/\gamma_{inclusive} + 1$ which is not exactly
equivalent. In contrast to the subtraction method where no
significant excess above the hadronic background could be seen for $p_{T} < 3$
GeV/$c$, a clear signal of direct photons is measured using the new
method, being about 10 \%. Both results are consistent.\\

The preliminary direct photon invariant yield for the 20 \% most central events
from the internal conversion measurement is shown in Figure
\ref{fig:auau_thermal} b) compared to different theoretical
calculations. The direct photon signal has been measured with large
significance for the shown $p_{T}$ range. The $T_{AB}$ scaled pQCD
calculation~\cite{Vogelsang} underpredicts the measurement at low
transverse momenta $p_{T} < 3$ GeV/$c$ while a 2+1 hydrodynamical
model for the emission of thermal photons with a formation time
$\tau_{0} = 0.15$ fm/$c$ and an average initial temperature
$T_{0}^{ave.} =$ 378 MeV
($T_{0}^{max} = 590$ MeV)~\cite{thermal} is significantly below the data at $p_{T} >
3$ GeV/$c$. A combination of both photons sources, however, describes
the data, but the temperature obtained for the thermal model has only
a meaning if the observed direct photon signal is of thermal
origin. Therefore it is necessary to measure the direct photons with
the same analysis technique in $p+p$ and $d+$Au collisions at the same
energy for $p_{T} < 3$ GeV/$c$ as well, where the excess will be much smaller if it is - in
Au+Au collisions - from thermal photons. The $d$+Au measurement will
also help understanding the influence of possible initial state
effects to the direct photon yield.

\section{Summary}
The PHENIX experiment has measured direct photons in $p+p$, $d$+Au and
Au+Au collisions at $\sqrt{s_{NN}}=200$ GeV. The measured yields in
all systems are consistent with a NLO pQCD calculation at high
transverse momenta ($p_{T} > 4.5$ GeV/$c$). Direct photons at $p_{T} <
4.5$ GeV/$c$ have been measured in Au+Au collisions via their internal
conversion into $e^{+}e^{-}$ pairs. A significant signal could be
extracted for $1 < p_{T} < 4.5$ GeV/$c$ that is significantly above
the expectation from NLO pQCD. However, also taking into account
thermal photon emissions, the measurement is in agreement with
calculations.

\end{document}